\magnification=\magstep1

\def\Sc{{\sl Science}}

\def\part{\partial}
\def\tsigma{{\tilde \sigma}}
\def\tr{{\rm tr}}
\def\frac#1#2{\hbox{$#1\over #2$}}
\def\B{{\cal B}}
\def\ii{\sqrt{-1}}

\baselineskip=24truept
\font\tif=cmr10 scaled \magstep3

\rightline{PUPT-1628}
\vfil
\centerline{\tif Quantum logic as a sum over classical logic gates}
\vfil
\centerline{
{\rm Bruno Nachtergaele}\footnote{${}^\ddagger$}{bxn@math.princeton.edu}
and {\rm Vipul 
Periwal}\footnote{${}^\dagger$}{vipul@puhep1.princeton.edu}
}
\bigskip
\centerline{Department of Physics}\centerline{Princeton University
}\centerline{Princeton, New Jersey 08544-0708}
\vfil
\par\noindent It is shown that certain natural quantum logic gates, {\it i.e.}
unitary time evolution matrices for spin-\frac{1}{2} quantum spins,
can be represented as sums, with appropriate phases,
 over classical logic gates, in a direct
analogy with the Feynman path integral representation of quantum 
mechanics.  On the other hand, it is shown that a natural quantum gate
obtained by analytically continuing the transfer matrix of the 
anisotropic nearest-neighbour Ising model to imaginary time, does not
admit such a representation.
\medskip
\vfil
\footline={\hfil}
\vfil\eject
\footline={\hss\tenrm\folio\hss}

\def\ANYAS{{\sl Ann.\ NY Acad.\ Sci.}}

\def\FP{{\sl Found.\ Phys.}}

\def\IJTP{{\sl Int.\ J. Theor.\ Phys.}}

\def\JSP{{\sl J. Stat.\ Phys.}}

\def\PTRSLA{{\sl Phil.\ Trans.\ Roy.\ Soc.\ Lond.\ A}}
\def\PD{{\sl Physica D}}

\def\PRA{{\sl Phys.\ Rev.\ A}}
\def\PRB{{\sl Phys.\ Rev.\ B}}

\def\PRL{{\sl Phys.\ Rev.\ Lett.}}
\def\PRSLA{{\sl Proc.\ Roy.\ Soc.\ Lond.\ A}}

\def\dajm{\hbox{D. A. Meyer}}

\def\feynman{\hbox{R. P. Feynman}}
\def\deutsch{\hbox{D. Deutsch}}

\def\shor{1}
\def\early{2}
\def\sig{3}
\def\bar{4}
\def\error{5}
\def\dec{6}
\def\path{7}
\def\paral{8}
\def\bruno{9}
\def\arley{11}
\def\cardy{10}

\hfuzz=3truept
Shor's discovery[\shor] of  polynomial time 
quantum algorithms for prime factorization and discrete logarithm has
resulted in an upsurge of interest in the properties of quantum 
computation[\early].  Significant results have been obtained concerning the
physical realizability of quantum gates, and the realizability of
classical universal 3-bit gates such as the Fredkin and Toffoli gates 
in terms of quantum 2-bit logic[\sig].  Furthermore,
Barenco et al.[\bar] have shown that all quantum gates can be expressed as 
compositions of all one-bit quantum gates and the two-bit exclusive-or gate.
The problems of error correction[\error] and 
decoherence[\dec] in quantum computation
have also been addressed.

Our aim here is to consider the properties of quantum logic in a 
different light.  Instead of constructing classical logic in terms of
quantum gates, we want to represent some rather general 
quantum gate arrays in terms of
coherent sums over classical gate arrays, much as Feynman represented quantum
mechanical amplitudes in terms of classical paths[\path].  There are obvious
reasons for wanting such a representation.  An intuition for
the efficiency of quantum computation is that quantum computers sum 
over many classical computations, and it is important to understand
quantitatively how this works, and how it can be exploited.   Further,
it is likely
that the true power of quantum computation will come from massively
parallel computation, a point that has been considered from the 
very beginnings of the subject, and recently re-emphasized[\paral].  
An intuition 
for the behaviour of such quantum logic, in terms of classical logic, 
is of great interest in this context, just as in the case of the Feynman
path integral.  
Consider, for example, the quantum phenomenon of tunneling---one would like
to know what the classical 
computational analogue of this might be, and how it should
be used in the design of quantum logic and quantum programming.  Indeed,
programming quantum logic on the basis of what classical logic it
replaces, is likely to be inefficient.  It may be more efficient to 
program according to the properties of quantum logic, much as with 
writing code for parallel processors, and for this purpose it is again 
important 
to gain some classical intuition for the properties of quantum gates.
Such classical logic representations also allow for a new type of 
simulation of quantum logic by classical parallel processors, rather
obviously.

We present two independent insights into classical representations of
quantum logic.  First, we show that for a natural set of Hamiltonians 
governing quantum spin-\frac{1}{2} degrees of freedom, there is a simple 
representation of the unitary time  evolution operator, in
other words, the quantum logic gate, in terms of appropriately
weighted sums over classical logic gates[\bruno].  We describe properties of
these `logic integrals' (adapting the term `path integrals' to the
present context) which can be deduced from the physical 
properties of the spin systems, and we suggest some uses for such
quantum logic. Certain general properties of such quantum logic for
a one-dimensional chain of spins could be 
inferred by finite size scaling calculations around conformal field
theories in two dimensions[\cardy].  

Secondly, we consider an anisotropic 
Ising model on a two-dimensional square lattice.
We show that the transfer matrix of this model, analytically continued,
is unitary at a unique value of the `time' coupling, and we show that 
this unitary quantum gate {\it cannot} be represented as a sum 
over classical logic gates in general.  Thus, `logic integrals' do not
necessarily exist as representations of quantum logic.  This will not
come as a surprise to physicists[\arley].

For our first problem, we consider quantum spin-$\frac 12$ degrees
of freedom  defined on a
finite set of sites $\Gamma.$ The Hilbert space at each site is
${\cal H}_x \cong {\bf C}^2,$ and observables are
elements of the bounded operators on this Hilbert space, just the set of
$2\times 2$ complex matrices $M(2,{\bf C}).$ The Hamiltonian $H$
for such a system can be written in general as
$$ H=-\sum_{b\in\B}J_b h_b$$
where  $\B,$ the set of `bonds', is a collection of
subsets of $\Gamma,$ and $h_b$ is an arbitrary element in 
$\otimes_{x\in b}M(2,{\bf C}).$   For much of our discussion, it will
suffice to take $\Gamma$ as a subset of the integers, say $\{0,\dots, L\},$
and $\B=\{\{0,1\},\dots,\{L-1,L\}\},$ which is the case easiest to
visualize, but it is important to observe that our approach holds in all
generality. 
Physically important observables are usually expressed in terms of
the spin matrices $S^1,S^2,S^3$ which are the generators of the
fundamental representation of SU(2).  They
satisfy the commutation relations 
$$ [S^\alpha,S^\beta]=\ii\sum_\gamma\epsilon_{\alpha\beta\gamma}S^\gamma $$
where $\alpha,\beta,\gamma\in\{1,2,3\}$ and
$\epsilon_{\alpha\beta\gamma}$ is the completely antisymmetric
tensor with $\epsilon_{123}=1$.

Aizenman and Nachtergaele[\bruno] have given a `quasi-state' decomposition
for the quantum statistical mechanics of this system, starting from
a Poisson integral formula. Using this decomposition for
`imaginary temperatures', we obtain the following expression
for the unitary evolution operator
$$\exp(\ii\beta H) = \int D_\beta\omega K(\omega),\eqno(1)$$
where $D_\beta\omega$, up to normalization and analytic
continuation, is the probability measure of a product of independent
Poisson processes for each bond in ${\cal B},$ running over the time interval
$[0,\beta]$ with rates $J_b$. More explicitly, the integration measure
is given by
$$D_\beta\omega = \prod_b\sum_{n_b=0}^\infty (-\ii J_b)^{ n_b}
\int_{0<t_i\le t_{i+1}\le \beta}\prod_{j=1}^{n_b} dt_j\eqno(2)$$
$ K(\omega)$ is a
time ordered product of operators, one for each bond in $\omega:$
$$K(\omega)={\prod}^* h_{b_n}h_{b_{n-1}}\dots h_{b_1},\eqno(3)$$
if $\omega$ is the set of bonds $\{(b_1,t_1),\dots, (b_n,t_n)\}$
with $t_1<t_2<\dots t_n.$ $\prod^*$ indicates a time ordered product.
Fig.~1 shows a configuration $\omega$ in the case that $\Gamma$ is 
a one-dimensional lattice and ${\cal B}$ are the nearest neighbour bonds
on this lattice.

Now consider, for example, $h$ to be the operator that interchanges the
states of the two sites, 
$$h\phi\otimes\psi=\psi\otimes\phi$$
 for any two vectors
$\phi,\psi \in {\bf C}^2.$  This is the exchange gate on 2 bits, 
$$E\equiv\left(\matrix{1&0&0&0\cr
0&0&1&0\cr
0&1&0&0\cr
0&0&0&1\cr}\right),$$
but it is equivalent to the Heisenberg spin $\frac 12$ ferromagnet! 
Fig.~2 shows a configuration $\omega$ in this model.

{From} the expressions (1-3) it is obvious that the quantum evolution 
operator can be decomposed as a linear superposition of classical logic gates
of the form $K(\omega)$. For concreteness, consider
a three spin system (equivalently, a three bit gate). By performing
the integrals in (2), we obtain series expansions
for the coefficients of the various classical logic gates appearing
in the decomposition:
$$\eqalign{\exp(\ii\beta H)= &(1-\beta^2 J^2+\dots){\bf 1} 
+ (-\ii \beta J +\dots)
E_{12} + (-\ii \beta J +\dots)E_{23} \cr &+ (- {1\over 2}\beta^2 J^2 +\dots)
E_{123} + (- {1\over 2}\beta^2 J^2 +\dots) E_{123}^2 + ({1\over 3}
\ii \beta^3 J^3 +\dots) E_{13}.\cr }$$
Here $E_{ij}$ is the exchange gate on the $i$ and $j$ bits, and 
$$E_{123} = E_{23}E_{12}=\left(\matrix{1&0&0&0&0&0&0&0\cr
0&0&0&0&1&0&0&0\cr0&1&0&0&0&0&0&0\cr0&0&0&0&0&1&0&0\cr
0&0&1&0&0&0&0&0\cr0&0&0&0&0&0&1&0\cr0&0&0&1&0&0&0&0\cr
0&0&0&0&0&0&0&1\cr}\right)$$
is the matrix that permutes the three bits cyclically. 
This example illustrates the utility of classical logic representations
of quantum gates---by varying $\beta,$ one can single out contributions
of different classical logic gates from the quantum gate.  $\beta$ is
just the time of evolution of the quantum system, so no external classical
`switches' are needed, which helps in minimizing the effects of 
decoherence[\dec].

If $\Gamma=\Gamma_A\cup\Gamma_B$ is a bipartite lattice, 
and ${\cal B}$  is a set with elements of the form $\{a,b\},a\in\Gamma_A$
and $b\in \Gamma_B,$ then we consider 
$$h=\sum_{m,m'=\pm 1/2} (-1)^{m-m'}|m,-m\rangle\langle m',-m'|.$$
This is the Heisenberg anti-ferromagnet.
In terms of classical logic, this operator $h$ corresponds to ${\bf 1}-E,$
and is shown in Fig.~3. In this case, a new phenomenon that contributes to
the quantum logic gate, 
but would not appear in classical logic, becomes apparent
in the quasi-state representation.  Notice  that $1-E$ is proportional to
a projection of rank 1: $(1-E)^2 = 2 (1-E).$  The factor of 2 corresponds 
to the fact that there are {\it closed loops} in a typical $\omega,$ as shown in
Fig.~3.
The sum over classical configurations that gives the quantum amplitude
therefore includes sums over `virtual' states of the classical logic.

Such logic integral decompositions of quantum logic can be extended to 
a much wider class of Hamiltonians with ease[\bruno], providing simple classical
logic representations with component classical gates that are $n$-bit
gates.  In the one-dimensional case, with $\Gamma=\{0,\dots,L\},$
this amounts to taking ${\cal B}=\{\{0,\dots,n\},\{1,\dots, n+1\},\dots\}.$  
Quasi-state decompositions for such cases have been 
reported in detail elsewhere[\bruno].

Simple properties of such massively parallel quantum logic can be extracted
from physical properties of these systems.  
Some quantum spin systems exhibit phase transitions at (imaginary)
values of $\beta$ in the infinite volume limit.  There are two aspects
of this that will be useful for quantum computation:
\item{1.} The existence of phase transitions implies that the quantum 
gate will exhibit different characteristics depending on the sign of
$\beta-\beta_c.$  In other words, letting the quantum evolution 
of the initial states run for a short time, or a long time, effectively
leads to two different quantum gates.  When $\beta > \beta_c,$ one 
expects long range order, implying algebraically decaying correlations
between the input and the output, and for $\beta<\beta_c$ one expects
exponential decay of correlations.
\item{2.} At finite lattice sizes, one can still get a good handle
on properties of the quantum gate for $\beta$ close to $\beta_c$
by calculating finite size corrections to the correlation functions 
at criticality[\cardy].

For the converse of our first problem, 
we turn now to the anisotropic Ising model in two dimensions, to exhibit
another aspect to representations of quantum logic as `logic
integrals' of classical logic.  Recall that this model
is a classical statistical mechanics model, with spins taking values
$\pm 1$ living on the sites of a square two dimensional lattice.  For
our purposes, we take the system to be of finite extent in the
space direction.  The time direction's extent will not be relevant
for us, but for the nonce we assume periodic boundary conditions in
the time direction.
The partition function of this model is 
$${\cal Z} \equiv \sum_{\{\sigma\}} \exp\left(-\beta_1 \sum_{i=0}^{N}\sum_{t}
\sigma_{i,t}\sigma_{i,t+1} -\beta \sum_{t}\sum_{i=0}^{N-1} 
\sigma_{i,t}\sigma_{i+1,t}\right),$$
where the sum over $t$ is a sum over the time slices of the lattice.
Introduce a transfer matrix $T,$ defined by
$$\langle \tsigma_{0},\dots \tsigma_{N}|T|\sigma_{0},\dots \sigma_{N}
\rangle \equiv 2^{-N/2}\exp\left(-\beta_1 \sum_{i=0}^{N}\tsigma_{i}\sigma_{i}
-\beta\sum_{i=0}^{N-1}\sigma_{i}\sigma_{i+1}\right).$$
For a lattice of time extent $\tau,$ the partition function 
can now be written as ${\cal Z} \propto \tr T^{\tau}.$  
This transfer matrix $T$ essentially allows one to interpret the Ising model
as a discrete-time one-dimensional quantum system, with $T\equiv
\exp(-H).$  We can now analytically continue this matrix to imaginary
time, and ask if there are imaginary values of $\beta$ and $\beta_1$ such that 
$T $ is a unitary matrix.   

To this end, we evaluate 
$$\langle \tsigma |TT^\dagger|\sigma\rangle
= 2^{-N}\sum_{\sigma'} \exp\left(- \sum_{i=0}^{N}[\beta_1\tsigma_{i}
+\beta_1^*\sigma_{i}] \sigma'_{i}
-(\beta+\beta^*)\sum_{i=0}^{N-1}\sigma'_{i}\sigma'_{i+1}\right).$$
It follows then that if $\beta=\sqrt{-1}\gamma,$ and $\beta_1 = 
\pm\sqrt{-1}{\pi/4},$ $T$ is a unitary matrix for any value of $\gamma.$

For $N=2,$ this matrix is
$$T= \left(\matrix{\ii&1&1&-\ii\cr
1&\ii&-\ii&1\cr
1&-\ii&\ii&1\cr
-\ii&1&1&\ii\cr}\right)\times {\rm diag}(\Delta,\Delta^*,\Delta^*,\Delta),$$
where $\Delta\equiv \exp(-\ii \gamma).$
If $\Delta=1,$ it is clear that $T$ can be written as
$$T(\gamma=0)=\ii\left[{\bf 1} - \ii\left(\matrix{0&1&0&0\cr 1&0&0&0\cr
0&0&0&1\cr0&0&1&0\cr}\right) - \ii\left(\matrix{0&0&1&0\cr0&0&0&1\cr1&0&0&0\cr
0&1&0&0\cr }\right) - \left(\matrix{0&0&0&1\cr0&0&1&0\cr0&1&0&0\cr1&0&0&0\cr}
\right)\right],$$
which is readily recognizable as a sum over classical logic gates, with
appropriate phase factors.  Here, 
we have restricted ourselves to decompositions
with coefficients of modulus 1.  It is easy to see that there are
two such decompositions.

However, when $\Delta\not=1,$ such a decomposition is not possible in
general.  Indeed, there is no reason to expect that it should be.  
Classical logic gates on $N$ bits are matrices in the
$2^N\times 2^N$ permutation representation of the permutation 
group on $2^N$ letters. By Schur's lemma, 
the complex linear span of the permutation representation on $n$ letters
is a strict subalgebra of the algebra of complex $n\times n$ matrices, since
the permutation 
representation is reducible, but the defining representation of U$(n)$ 
is certainly irreducible, so its complex linear span is all of the 
algebra of complex $n\times n$ matrices. 

In conclusion, we have shown that there is a natural 
representation of parallel quantum gates in terms of sums over
classical logic gates, analogous to the Feynman sum over paths
representation of quantum mechanical amplitudes.  We have shown that
this viewpoint on quantum logic allows a whole host of tools from
statistical mechanics and quantum spin chains to be used to obtain a
better intuition for the characteristics of quantum logic.  We have 
explicitly shown that such representations may not always be possible,
indicating some of the limits of this approach.

\centerline{References}
\def\hfb{\enspace \relax}
\medskip
\item{\shor} P. W. Shor,
``Algorithms for quantum computation:  discrete logarithms and
  factoring'',
in S. Goldwasser, ed.,
{\sl Proc. 35th Symp. on Foundations of Computer
Science}, Santa Fe, NM, 20--22 November 1994
(Los Alamitos, CA:  IEEE Computer Society Press 1994) 124--134

\item{\early} P. Benioff, {\sl J. Stat. Phys.} {\bf 22} (1980) 563; {\bf 29}
(1982) 515; \feynman,
``Simulating physics with computers'',
\IJTP\ {\bf 21} (1982) 467--488;
``Quantum mechanical computers'',
\FP\ {\bf 16} (1986) 507--531; 
\deutsch,
``Quantum theory, the Church--Turing principle and the universal
  quantum computer'',
\PRSLA\ {\bf 400} (1985) 97--117;
A. Peres, ``Reversible logic and quantum computation'',
{\sl Phys. Rev.} {\bf A32} (1985) 3266;
\deutsch\ and R. Jozsa,
``Rapid solution of problems by quantum computation'',
\PRSLA\ {\bf 439} (1992) 553--558;\hfb
E. Bernstein and U. Vazirani,
``Quantum complexity theory'',
in {\sl Proc. 25th ACM Symp. on Theory of Computing},
San Diego, CA, 16--18 May 1993
(New York:  ACM Press 1993) 11--20
A. Berthiaume and G. Brassard,
``The quantum challenge to structural complexity theory'',
in {\sl Proc. 7th Structure in Complexity Theory
Conference}, Boston, MA, 22--25 June 1992
(Los Alamitos, CA:  IEEE Computer Society Press 1992) 132--137;\hfb
D. R. Simon,
``On the power of quantum computation'',
in S. Goldwasser, ed.,
{\sl Proc. 35th Symp. on Foundations of Computer
Science}, Santa Fe, NM, 20--22 November 1994
(Los Alamitos, CA:  IEEE Computer Society Press 1994) 116--123

\item{\sig}  S. Lloyd, ``A potentially realizable quantum supercomputer'',
{\sl Science} {\bf 261} (1993) 1569; R. Landauer,
``Is quantum mechanics useful?'',
\PTRSLA\ {\bf 353} (1995) 367--376;
M. B. Plenio and P. L. Knight,
``Realistic lower bounds for the factorization time of large numbers
  on a quantum computer'',
preprint (1995) quant-ph/9512001;\hfb
D. Beckman, A. N. Chari, S. Devabhaktuni and J. Preskill,
``Efficient networks for quantum factoring'',
preprint (1996) CALT-68-2021, quant-ph/9602016; 
N. Margolus,
``Quantum computation'',
\ANYAS\ {\bf 480} (1986) 487--497; \dajm,
``From quantum cellular automata to quantum lattice gases'',
UCSD preprint (1995), quant-ph/9604003, to appear in \JSP;
I. L. Chuang and Y. Yamamoto,
``A simple quantum computer'',
\PRA\ {\bf 52} (1995) 3489--3496;
T. Sleator and H. Weinfurter, ``Realizable Universal Quantum Logic
Gates'', \PRL\ {\bf 74} (1995) 4087--4090;
H. K\"orner and G. Mahler,
``Optically driven quantum networks:  applications in molecular
  electronics'',
\PRB\ {\bf 48} (1993) 2335--2346;
A. Barenco, D. Deutsch, A. Ekert and R. Jozsa,
``Conditional quantum dynamics and logic gates'',
\PRL\ {\bf 74} (1995) 4083-4086;\hfb
J. I. Cirac and P. Zoller,
``Quantum computations with cold trapped ions'',
\PRL\ {\bf 74} (1995) 4091--4094;
C. Monroe, D.M. Meekhof, B.E. King, W.M. Itano and D.J. Wineland,
``Demonstration of a universal quantum logic gate'', \PRL {\bf 75} (1995) 4714;
W.H. Zurek and R. Laflamme, ``Quantum Logical Operations on Encoded Qubits'',
quant-ph/9605013;
D. P. DiVincenzo,
``Two-bit gates are universal for quantum computation'',
\PRA\ {\bf 51} (1995) 1015--1022;\hfb
H.F. Chau and F. Wilczek, ``Realization of the Fredkin gate using a series
of one- and two-body operators, quant-ph/9503005 

\item{\bar}A. Barenco, C. H. Bennett, R. Cleve, D. P. DiVincenzo, N. Margolus,
P. Shor, T. Sleator, J. Smolin and H. Weinfurter,
``Elementary gates for quantum computation'',
\PRA\ {\bf 52} (1995) 3457--3467;\hfb

\item{\error} I.L. Chuang and R. Laflamme, ``Quantum Error Correction by 
Coding'', quant--ph/9511003; R. Laflamme, C. Miquel, J.P. Paz and W.H. Zurek,
``Perfect Quantum Error Correction Code'', quant-ph/9602019;
E. Knill and R. Laflamme, ``A Theory of Quantum Error-Correcting Codes'',
quant-ph/9604034; 
 A.R. Calderbank and P.W. Shor, ``Good quantum error 
correcting codes exist'', quant-ph/9512032;
P.W. Shor and J.A. Smolin, ``Quantum Error Correcting
Codes Need Not Completely Reveal the Error Syndrome'', quant-ph/9604006;
A.M. Steane, ``Multiple Particle Interference and Quantum Error Correction'',
quant-ph/9601029; ``Simple Quantum Error Correcting Codes'',  quant-ph/9605021

\item{\dec} W. G. Unruh,
``Maintaining coherence in quantum computers'',
\PRA\ {\bf 51} (1995) 992--997;\hfb
P.W. Shor, ``Scheme for reducing decoherence in quantum memory'', \PRA 
{\bf 52} (1995) 2493; 
G. M. Palma, K.-A. Souminen and A. Ekert,
``Quantum computers and dissipation'',
\PRSLA\ {\bf 452} (1996) 567--584;
I. L. Chuang, R. Laflamme, P. Shor and W. H. Zurek,
``Quantum computers, factoring and decoherence'',
\Sc\ {\bf 270} (1995) 1633-1635;
C. Miquel, J. P. Paz and R. Perazzo,
``Factoring in a dissipative quantum computer'',
preprint (1996) quant-ph/9601021; 
 R.J. Hughes, D.F.V. James, E.H. Knill, R. Laflamme, A.G. Petschek,
``Decoherence Bounds on Quantum Computation with Trapped Ions'',
quant-ph/9604026; I.L. Chuang, R. Laflamme and J.P. Paz, ``Effects of Loss 
and Decoherence on a Simple Quantum Computer'', quant-ph/9602018

\item{\path}
\feynman\ and A. R. Hibbs,
{\sl Quantum Mechanics and Path Integrals}
(New York:  McGraw-Hill 1965)

\item{\paral} R.P. Feynman, Caltech Physics X lectures (1981);  W.D. Hillis,
``New computer architectures and their relationship to physics or
  why computer science is no good'',
\IJTP\ {\bf 21} (1982) 255--262;\hfb N. Margolus,
``Parallel quantum computation'',
in W. H. Zurek, ed.,
{\sl Complexity, Entropy, and the Physics of Information},
proceedings of the SFI Workshop, Santa Fe, NM,
29 May--10 June 1989,
{\sl SFI Studies in the Sciences of Complexity} {\bf VIII}
(Redwood City, CA:  Addison-Wesley 1990) 273--287;\hfb
N. Margolus,
``Physics-like models of computation'',
\PD\ {\bf 10} (1984) 81--95;
R. Mainieri,
``Design constraints for nanometer scale quantum computers'',
preprint (1993) LA-UR 93-4333, cond-mat/9410109

\item{\bruno} M. Aizenman and B. Nachtergaele, ``Geometric Aspects of
Quantum Spin States'', {\sl Comm. Math. Phys.} {\bf 164} (1994) 17--63;
B. Nachtergaele, ``Quasi-state decompositions for quantum spin systems'',
{\sl Prob. Theory and Math. Stat.}, Proc. Sixth Vilmius Conference,
(eds. B. Grigelionis et al.) (Publishing Services Group, Vilnius, 1994)

\item{\cardy} See, for example, 
J. Cardy, ``Critical Percolation in Finite Geometries'',
{\sl J. Phys.} {\bf A25} (1992) L201--L206

\item{\arley} A recent discussion of the sum over paths can be found in
A. Anderson, ``The use of exp$(iS[x])$ in the sum over histories'', 
{\sl Phys. Rev.} {\bf D49} (1994) 4049, and references therein

\vfill\eject
\nopagenumbers
\centerline{Figure Captions}
\bigskip
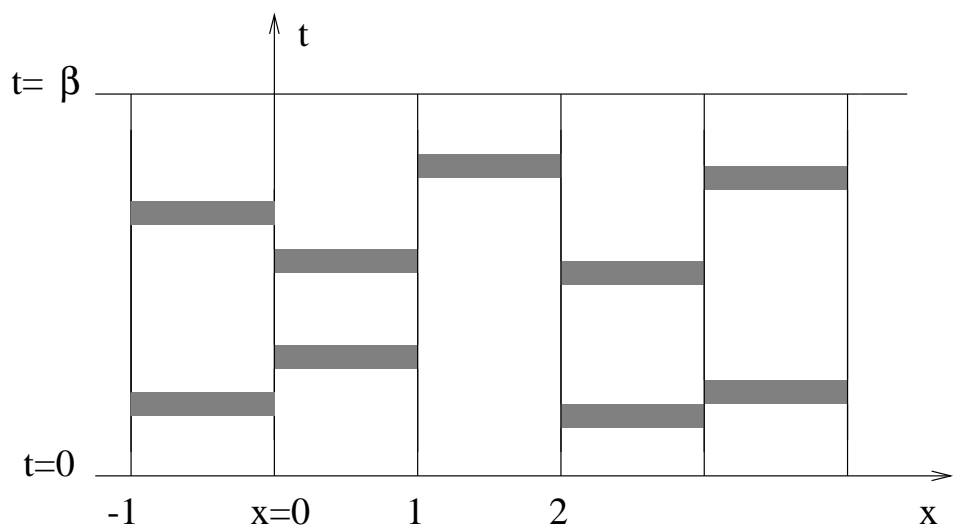
\item{Fig. 1} A space-time configuration $\omega$ for a general Hamiltonian
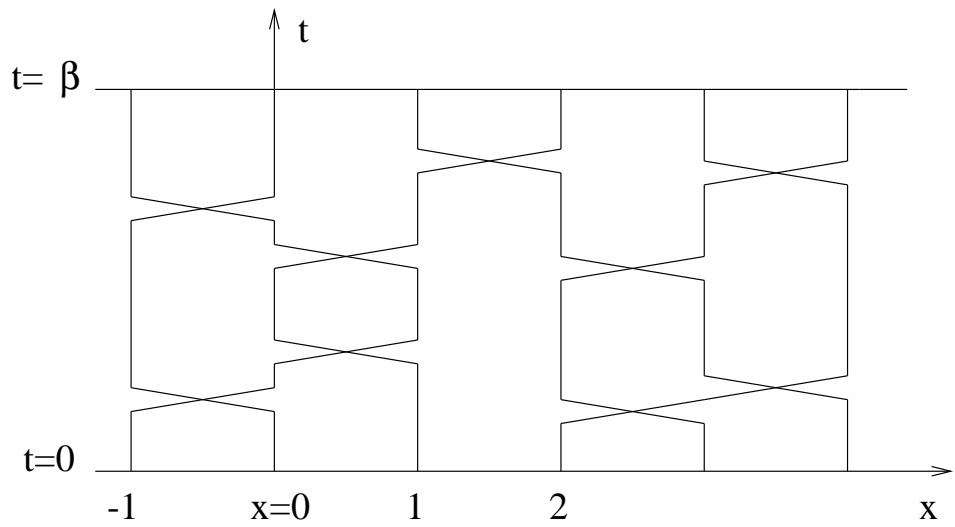
\item{Fig. 2} A configuration $\omega$ for the one-dimensional Heisenberg
spin-\frac12 ferromagnet
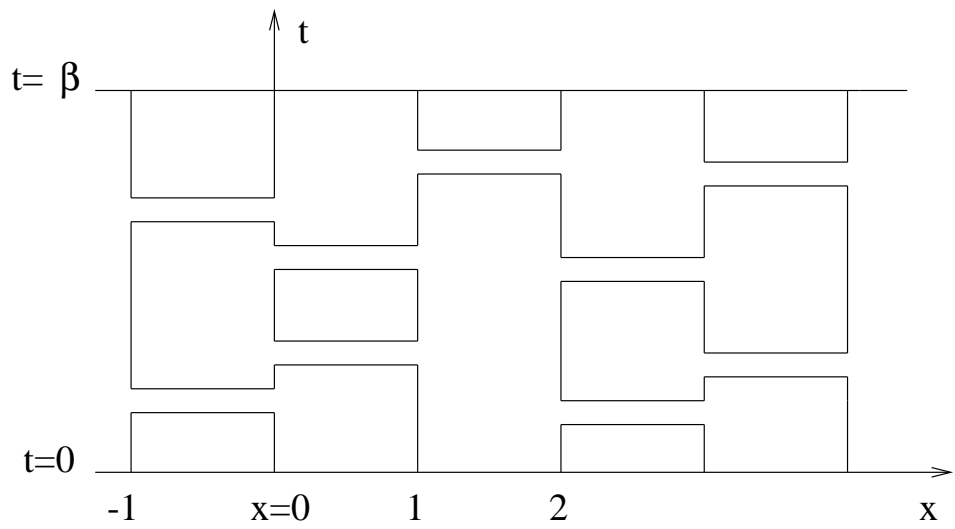
\item{Fig. 3} A configuration $\omega$ for the one-dimensional Heisenberg
antiferromagnet, showing virtual loops

\end